\documentclass[runningheads]{llncs}
\usepackage[T1]{fontenc}
\usepackage{makecell}
\usepackage{ulem}
\usepackage{amsmath}
\usepackage{amsfonts}
\usepackage{booktabs}
\usepackage{hyperref}
\usepackage{color}
\usepackage{graphicx}
\usepackage{cleveref}

\begin{document}
\title{Improving Audio Caption Fluency with Automatic Error Correction}

\author{Hanxue Zhang*
\and
Zeyu Xie*
\and
Xuenan Xu
\and
Mengyue Wu$\dag$
\and
Kai Yu$\dag$\thanks{*Equal contribution; $\dag$ Corresponding authors.}}
\authorrunning{H. Zhang et al.}
\institute{MoE Key Lab of Artificial Intelligence\\
  X-LANCE Lab, Department of Computer Science and Engineering\\
  AI Institute, Shanghai Jiao Tong University, Shanghai, China\\}
%\email{\{zhx_jiaxue,zeyu_xie\}@sjtu.edu.cn}
%
\maketitle              % typeset the header of the contribution
\begin{abstract}
Automated audio captioning (AAC) is an important cross-modality translation task, aiming at generating descriptions for audio clips.
However, captions generated by previous AAC models have faced ``false-repetition'' errors due to the training objective.
% In such a case, we propose a new task called Audio Captioning Correction (ACC) and hope to reduce such errors by secondary processing of the results.
In such scenarios, we propose a new task of AAC error correction and hope to reduce such errors by post-processing AAC outputs.
% To tackle this problem, we present a dataset which includes pseudo sentences with ``false-repetition'' errors generated by rules.
To tackle this problem, we use observation-based rules to corrupt captions without errors, for pseudo grammatically-erroneous sentence generation.
One pair of corrupted and clean sentences can thus be used for training.
We train a neural network-based model on the synthetic error dataset and apply the model to correct real errors in AAC outputs.
Results on two benchmark datasets indicate that our approach significantly improves fluency while maintaining semantic information.

\keywords{Error correction  \and Audio captioning \and False-repetition \and Long short-term memory}
\end{abstract}
\section{Introduction}
Automated audio captioning (AAC) is a cross-modal translation task from audio to natural language description~\cite{drossos2017automated,wu2019audio}.
Taking audio as the input and text as the output, AAC is similar to Automated Speech Recognition (ASR) but does not require alignment between source audio and target text sequence.
Compared with other audio understanding tasks such as Acoustic Scene Classification (ASC) and Sound Event Detection (SED), AAC focuses on unstructured free text but not pre-defined classes, allowing more flexible, multi-faceted description. 
Potential applications of AAC include retrieving multimedia content by audio and analyzing sounds for security surveillance.

Recently, several approaches are proposed to improve AAC performance.
% Pre-trained audio classification and text generation models are transferred to AAC~\cite{koizumi2020audio,mei2021audio,xu2021investigating} to alleviate the data scarcity problem.
% A few works explore the incorporation of semantic clue~\cite{eren2020semantic,gontier2021automated,koizumi2020transformer} to provide guidance for caption generation.
Despite the remarkable performance improvement achieved, we find that grammatical errors are commonly observed in captions generated by these methods.
\Cref{tab:grammar_error_summary} summarizes some of the main error types and their corresponding examples.
The majority of these errors are not caused by the misunderstanding of the audio content, but by the text generation process.
For example, some models tend to contain repeated events in a single caption, which is called ``false-repetition'' error.
We assume that such false-repetition errors can be corrected with only captions as the input, without the reference of audio clips.
In other words, grammatical errors such as missing sound events are not included in false-repetition errors since correcting these errors require audio clip reference.
% Although grammatical errors are common in model outputs, the problem attracts little attention in previous works.
% The only related work is a newly-proposed metric FENSE~\cite{zhou2022can} which attaches importance to penalizing captions with grammatical errors during evaluation.

% More disturbingly, the widely used evaluation metrics (such as BLUE) are not sensitive to such repetition errors~\cite{zhou2022can}.
% Accordingly, correcting the "false-repetition" flaw of the caption model while preserving the original audio information, remains a challenging task. 

\begin{table}
\centering
\caption{The main error types and corresponding examples of the ``false-repetition'' issue.}
\label{tab:grammar_error_summary}
\begin{tabular}{c|c}
\toprule
\textbf{Error Types} & \textbf{Example} \\
\midrule
Verb Repetition & A bird is singing \textbf{and singing} and chirping. \\
\hline
Sentence Repetition & A dog is barking \textbf{and a dog is barking}. \\
\hline
Partial Repetition & A woman speaks and music plays \textbf{as a woman speaks}. \\
\hline
Adverb Repetition & A bird is chirping loudly \textbf{loudly}.    \\
\hline
Extra Tails & A door is being opened and closed \textbf{and then}. \\
%\hline
%TODO xnx is this false repetition?
%Missing Component & A \textbf{(man / woman / $\cdots$)} is talking in the background.  \\
\bottomrule
\end{tabular}
\end{table}

To alleviate the aforementioned grammatical problems in AAC output, we propose a neural network-based approach to automatically correct false-repetition errors.
We synthesize parallel training data by corrupting ``clean'' sentences, i.e., sentences without grammatical errors.
Reference captions from two benchmark datasets, Clotho~\cite{drossos2020clotho} and AudioCaps~\cite{kim2019audiocaps}, are taken as clean sentences and corrupted by rules to generate parallel training pairs.
We adopt a bidirectional long short-term memory BiLSTM~\cite{zhang2015bidirectional} network as the error correction model to discriminate whether each word of the generated caption should be deleted.
The model is trained on synthesized training pairs.
After training on synthesized sentence pairs, we apply the model to AAC model outputs. 
Results demonstrate that the model effectively increases the fluency of AAC outputs by reducing false-repetition errors.

\section{Related Work}
\subsection{AAC Work}
AAC task has undergone significant development recently. 
Several works introduced audio encoders and text generation decoders pre-trained on large-scale data~\cite{koizumi2020audio,mei2021audio,xu2021investigating}.%to enhance the model's ability to understand audio clip and to output sentences with relevant semantic content.
Pre-training alleviates the problem of the small size of available AAC training data.
Auxiliary semantic information has also been introduced in some studies to improve the description accuracy~\cite{eren2020semantic,gontier2021automated,koizumi2020transformer}.
Reinforcement learning is used to directly improve specific metrics~\cite{xu2020crnn}, such as CIDEr~\cite{vedantam2015cider}.
% For some metrics of AAC tasks, such as CIDER, the reinforcement learning method may improve the performance of the model involving these objective evaluation system.

However, the works mentioned above pay more attention to enhancing objective metrics' performance by generating more accurate and detailed audio description, while ignoring the grammatical error in generated captions.
Captions generated by an expected system should be not only accurate and detailed, but also \textit{fluent}.
FENSE~\cite{zhou2022can}, a recently proposed evaluation metric, takes fluency into consideration and incorporates grammatical error detection to recognize erroneous sentences and penalize them accordingly. %so that evaluate audio descriptions more comprehensively. 
Repetitive (or extra) and missing elements are detected as grammatical errors in FENSE.
Inspired by FENSE, we aim to reduce errors of repetitive elements by training an automatic correction model.
% However, Fense only made error detection and did not involve the method of sentence correction.

%Is there any work on AAC models themselves to reduce false-repetition errors? i don't know - 

%fense
%\MYW{you introduce previous (related) aac work and the kind of grammatical errors existed under different models}

\subsection{Error Correction}
Grammatical Error Correction (GEC) was first proposed in Natural Language Processing (NLP).
There are many studies on pseudo data generation.
Lichtarge et al.~\cite{lichtarge2019corpora} accumulated source-target pairs from the Wikipedia revision histories.
Wan et al.\cite{wan2020improving} generated synthetic pairs by editing latent representations of grammatical sentences.
Researchers applied several frameworks to grammatical error correction.
Malmi et al.~\cite{malmi2019encode} proposed an approach to transforming error correction to a text editing task.
% Chen et al.~\cite{chen2020improving} feed sentences to the seq2seq model for correction. 

Recently, error correction techniques have been vastly adopted in ASR post-processing.
Tanaka et al.~\cite{tanaka2018neural} trained a language model to choose better results among ASR output candidates.
Mani et al.~\cite{mani2020asr} proposed a Transformer-based model to correct ASR outputs in an auto-regressive way.
Liao et al.~\cite{liao2020improving} further combined MASS~\cite{song2019mass} pre-training with ASR correction.
%TODO xnx what types of error in GEC and ASR? delete? insertion? substitution?

To the best of our knowledge, there are no prior works on error correction of AAC outputs.
% Inspired by GEC and ASR correction tasks, we creatively proposed to transfer error correction models to AAC tasks.
Error types in GEC and ASR often involve only small parts of the sentence where substitution and insertion are inevitable in correction.
However, we focus on false-repetition errors where only deletion is needed.
Due to the difference between error patterns, we do not use common approaches in GEC or ASR correction such as edit distance~\cite{leng2021fastcorrect} or sequence to sequence (S2S) models~\cite{chen2020improving}.
Instead, we treat AAC error correction as a sequence labeling task to decide which words in the generated caption should be deleted.

\begin{table}
\centering
\caption{Pseudo data generation upon observation-based rules}
\label{tab:data_generation_rules}
\begin{tabular}{l|l}

\toprule
\textbf{Error Types} & \textbf{Rules} \\
\midrule
Verb Repetition & {\makecell[l]{Repeat a verb with a conjunction.\\
e.g., A baby is crying \sout{and crying}}}\\
\hline
Adverb Repetition & {\makecell[l]{Repeat the adverb . \\
e.g., Baby cries loudly \sout{loudly}}}\\
\hline
Partial Repetition & {\makecell[l]{ Repeat part of the sentence.\\
e.g., A man talks followed by a cat meows \sout{and a cat meows}}}\\
\hline
Sentence Repetition & {\makecell[l]{ Repeat the whole sentence with conjunctions.\\
e.g., A train passed by \sout{while a train passed by}}}\\
\hline
Extra tails & {\makecell[l]{Add useless tails to the end of the sentence.\\
e.g., The audience laughed and applauded \sout{and a}}}\\
\bottomrule
\end{tabular}
\end{table}

\section{AAC Automatic Error Correction}
Our proposed method consists of two phases: pseudo-error data generation and correction model training, respectively introduced in the following sections.  
\subsection{Pseudo Data Generation}

\begin{figure}[htbp]
\centering
\begin{minipage}[t]{0.48\linewidth}
\centering
\includegraphics[width=\linewidth]{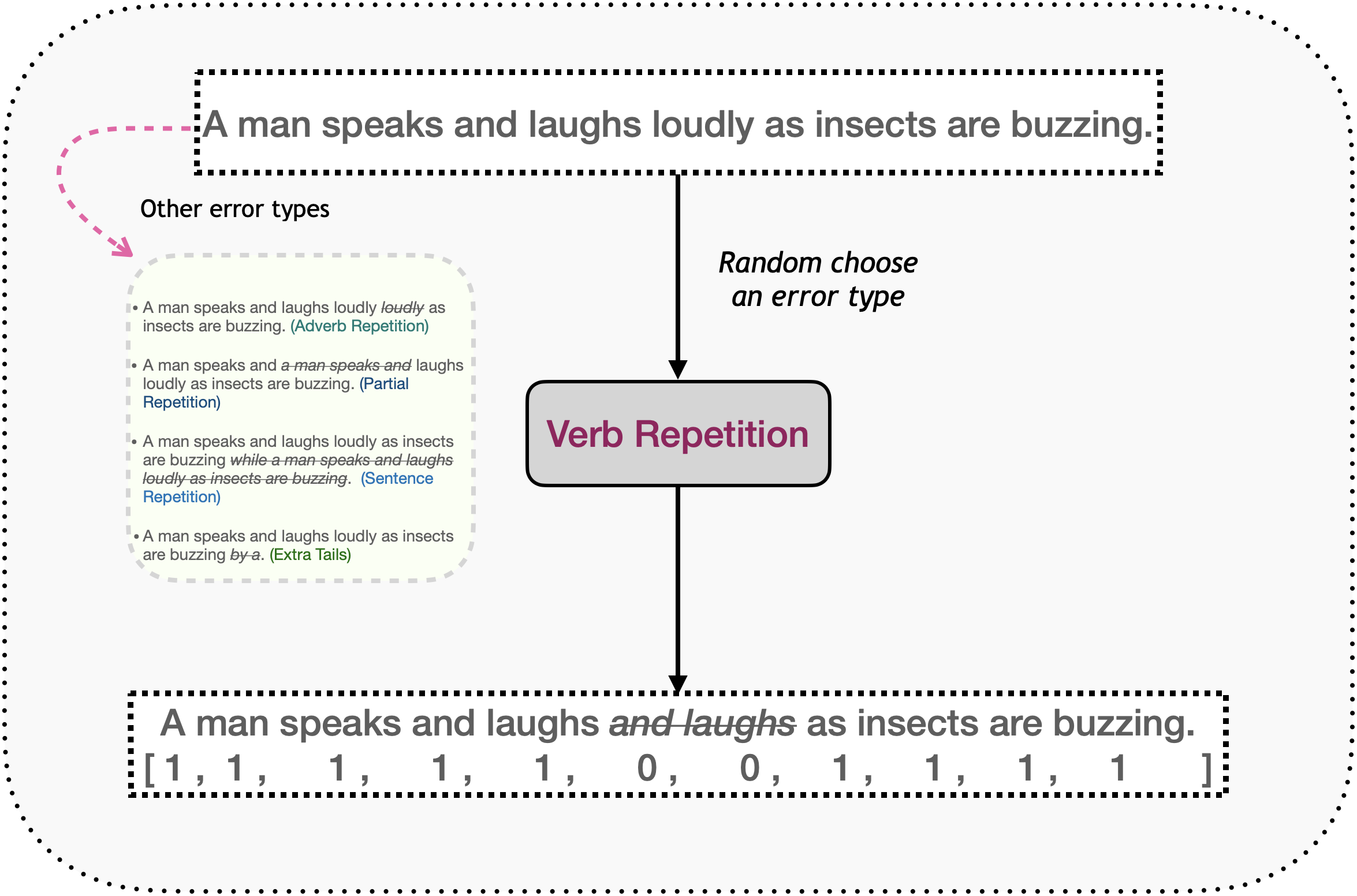}
\caption{An example to generate pseudo\\ sentence.}
\label{fig:data_generation}
\end{minipage}%
\quad
\begin{minipage}[t]{0.48\linewidth}
\centering
\includegraphics[width=0.98\linewidth]{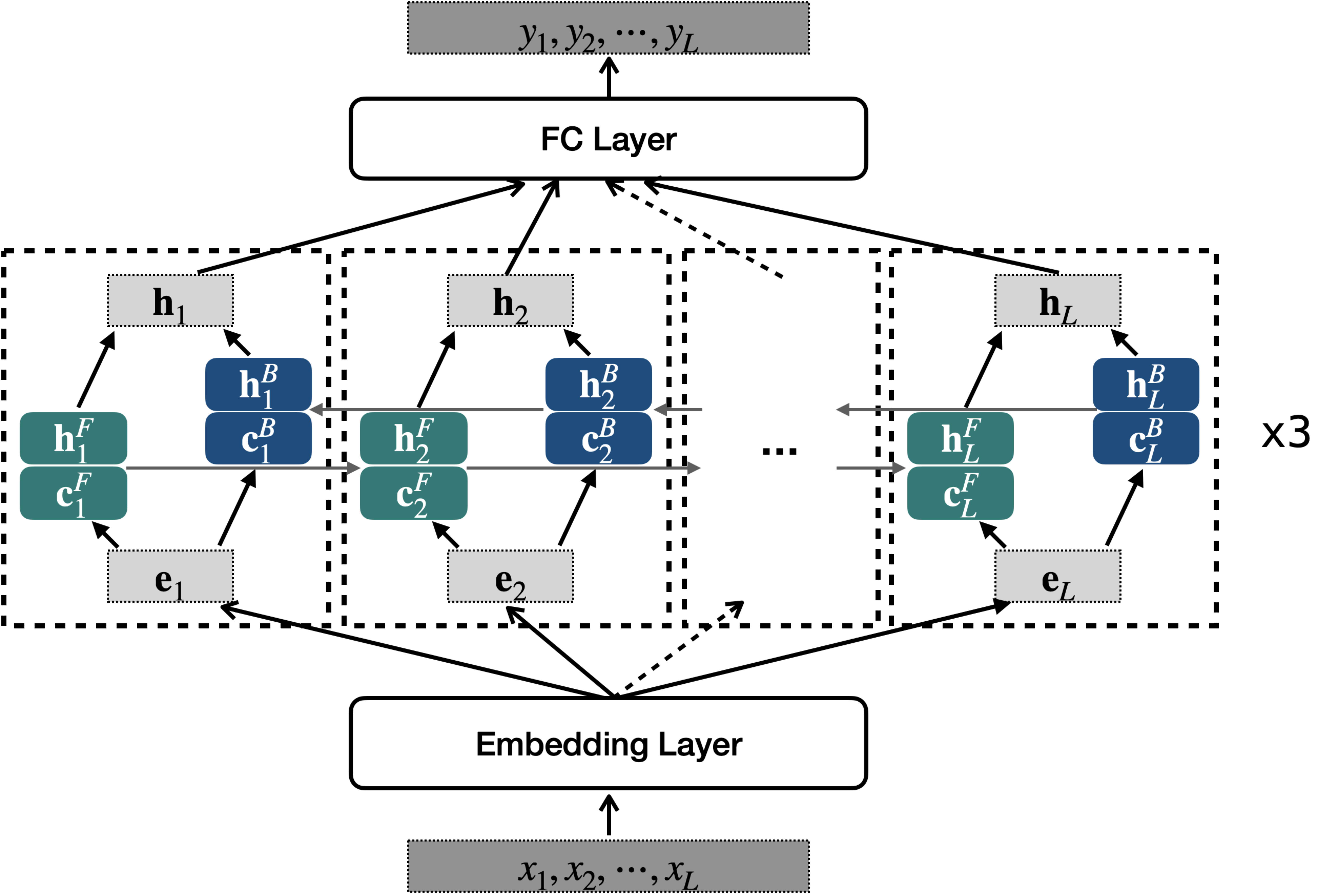}
\caption{The three-layer BiLSTM error correction model structure.}
\label{fig:model}
\end{minipage}
\end{figure}

Due to the uniqueness of error patterns in AAC outputs, sentence pairs with real-world grammatical errors are unsuitable for AAC correction training.
We generate pseudo grammatically-erroneous sentences by corrupting sentences with rules summarized in \Cref{tab:data_generation_rules}, which are defined according to the corresponding error types in \Cref{tab:grammar_error_summary}. 
Clean sentences are collected from AudioCaps and Clotho references.
Grammatically-erroneous sentences are generated by repeating some components of clean sentences or adding extra tails.
For a sentence (whether clean or corrupted) with $L$ words, its label sequence $\mathbf{y}=\{y_i\}_{i=1}^L,y_i\in\{0, 1\}$ mark grammatically-erroneous parts with 0.
In other words, words with the label 0 should be deleted.
Clean captions would be labeled with $\{1\}^L$.
\Cref{fig:data_generation} gives an example.
The dataset contains both ``corrupted-clean'' and ``clean-clean'' sentence pairs to prevent the model from deleting correct parts of the input sentence.

\subsection{Correction Model}
The proposed model is shown in \Cref{fig:model}, which consists of a word embedding layer, a three-layer BiLSTM and a fully connected (FC) output layer.
Formally, given the input text sequence $\mathbf{x}=\{x_i\}_{i=1}^{L}$, with the length $L$:
\begin{align}
    \begin{split}
        \mathbf{x}^{E} &= \mathrm{WE}(\mathbf{x})\\
        \mathbf{h}_{i}^{1} &= \mathrm{BiLSTM}(\mathbf{x}_{i}^{E}, \mathbf{h}_{i-1}^{1})\\
        \mathbf{h}_{i}^{l+1} &= \mathrm{BiLSTM}(\mathbf{h}_{i}^{l}, \mathbf{h}_{i-1}^{l+1}), \quad l = 1, 2\\
        \mathbf{y} &= \mathrm{FC}(\mathbf{h}^{3})
    \end{split}
\end{align}

where WE indicates the word embedding layer which maps the discrete text sequence into continuous space.
BiLSTM encodes the text sequence forward and backward word by word recurrently. 
An FC layer followed by a softmax function works as a classifier to output $\mathbf{\hat{y}}$.
The model is trained by cross-entropy loss.
After training, the correction model is used to correct false-repetition errors in captions.
The model output $\mathbf{\hat{y}}$ is used as a mask where 0 denotes words to be deleted in the input sentence.

\section{Experiments Setup}

\subsection{Dataset}
Two benchmark AAC datasets, Clotho and AudioCaps are used in our experiments.
AudioCaps is the largest public dataset for AAC at present. 
The audio clips are selected from the large-scale manually labeled audio event dataset AudioSet.
It includes about $50k$ audio clips, each of which lasts no more than $10 $ seconds. 
AudioCaps provides one annotation for each audio clip in the training set while $5$ in the validation and test sets.
Clotho includes $5929$ audio clips, each of which is $15$ to $30$ seconds and provided with $5$ annotation captions.
Clotho is used for the AAC task in the Detection and Classification of Acoustic Scenes and Events (DCASE) challenges in recent years.

% We exclude the impact of capitalization and punctuation and create a vocabulary to represent the word discretely with the index in the vocabulary.
%TODO xnx check the rule
For each reference sentence in both AudioCaps and Clotho, we generate five corrupted captions, each containing one error type in \Cref{tab:data_generation_rules}.
Along with the original correct captions, the generated dataset is divided into training, validation and test sets. %which maintain the same distribution as the original AudioCaps and Clotho.

\subsection{AAC Model}
The baseline AAC system consists of an audio encoder and a text decoder \cite{xu2022sjtu}.
%TODO xnx give a reference and describe less
The audio encoder is a three-layer GRU encoder for both AudioCaps and Clotho.
Regarding the text decoder, a bidirectional GRU (BiGRU) is used for Clotho while a Transformer is used for AudioCaps.

Based on the baseline system, reinforcement learning (RL) is further applied.
It is reported that models trained by RL are more prone to generate grammatically-erroneous captions though evaluation metrics are improved~\cite{mei2021encoder}.
%on both AudioCaps and Clotho since RL is usually more likely to cause grammatical errors while improving specific metrics.

\subsection{Correction Model Configuration}

% The total vocabulary size $6425$ is set as the size of the word embedding layer. 
Hidden dimension is set to $256$ in BiLSTM with a dropout of $0.5$.
The whole model is trained in $25$ epoch with Adam optimizer.
The learning rate is set to $1e^{-6}$ at beginning and decreases exponentially to $5e^{-7}$.

After training, the correction system is used to delete false-repetition parts in AAC outputs.
% Based on the output of AAC baseline model and RL model, the correction system uses the output mask to correct grammatical errors.
The outputs before and after correction are evaluated in terms of $\mathrm{BLEU}_\mathrm{4}$~\cite{kishore2002bleu}, $\mathrm{ROUGE}_\mathrm{L}$~\cite{chin-yew2004rouge}, METEOR~\cite{lavie2007meteor}, CIDEr~\cite{vedantam2015cider}, SPICE~\cite{anderson2016spice} and FENSE.

\subsection{Results}

\begin{table}[htpb]
 
  \caption{Results of system performance. ER indicates error correction processing using the correction model. Results involving ER are the average results of four random seeds. RL indicates systems trained using reinforcement learning. $\mathrm{FENSE}_\mathrm{w/o \, penalty}$ indicates that error detector and penalization are disabled in FENSE.}
  \label{tab:caption_result}
  \centering
  \begin{tabular}{c|ccccccc}
    \toprule
    System & $\mathrm{BLEU}_\mathrm{4}$ & $\mathrm{ROUGE}_\mathrm{L}$ & METEOR & CIDEr & SPICE & FENSE & $\mathrm{FENSE}_\mathrm{w/o \, penalty}$\\
    \hline
    \hline
    & \multicolumn{7}{c}{Clotho}\\
    \hline
     base
    &16.4 &38.6 &18.1 &42.1 &12.6 &45.0	&49.8\\

    base + ER
    & 16.9&38.8&18.0&42.9&12.6&\textbf{46.4}&49.3\\  %16.9($\pm$0.0)&38.8($\pm$0.0)&18.0($\pm$0.0)&42.9($\pm$0.0)&12.6($\pm$0.0)&46.4($\pm$0.0)&49.3($\pm$0.0)
    \hline
    RL
    & 17.2 &40.9 &18.1 &48.7 &12.1 &20.7 &49.1\\

    RL + ER
    & 17.4&39.8&17.5&45.9&12.1&\textbf{30.8}&49.2\\    %17.4($\pm$0.0)&39.8($\pm$0.1)&17.5($\pm$0.0)&45.9($\pm$0.7)&12.1($\pm$0.0)&30.8($\pm$9.5)&49.2($\pm$0.0)\\
    
    \hline
    \hline
    & \multicolumn{7}{c}{AudioCaps}\\
    \hline
    base
    &28.2 &50.1 &25.0 &73.4 &18.6 &61.6 &63.0\\

    base + ER
    & %28.5($\pm$0.0)&50.0($\pm$0.0)&25.0($\pm$0.0)&73.8($\pm$0.0)&18.6($\pm$0.0)&62.0($\pm$0.0)&63.0($\pm$0.0)\\
    28.5&50.0&25.0&73.8&18.6&\textbf{62.0}&63.0\\
    \hline
    
     RL
    &27.5 &50.0 &24.5 &84.1 &16.3 &21.8 &62.9\\

    RL + ER
    & %27.8($\pm$0.0)&49.0($\pm$0.0)&24.0($\pm$0.0)&82.2($\pm$0.0)&17.8($\pm$0.0)&45.8($\pm$0.1)&63.0($\pm$0.0)\\
    27.8&49.0&24.0&82.2&17.8&\textbf{45.8}&63.0\\
    
    \bottomrule
  \end{tabular}
\end{table}

After training on the generated error data, the BiLSTM model achieves an accuracy of 99.9 and an macro-F1 score of 99.8. 
We further utilize the model to conduct post-processing on audio captioning model outputs.
The performance of the proposed method is shown in \Cref{tab:caption_result}.
Metrics except for FENSE measure the semantic similarity between generated captions and references.
FENSE combines semantic similarity and grammatical error penalty, which presents a higher correlation with human judgments.
Among these metrics, FENSE significantly improves, which shows that the correction model effectively reduces grammatical errors in captioning outputs by post-processing, so that sentences are much less penalized.
For contrast, we use $\mathrm{FENSE}_\mathrm{w/o \, penalty}$ to evaluate the semantic accuracy of generated captions.
% It is more reliable by using SentenceBERT~\cite{reimers2019sentence} for evaluation.
% \MYW{Introducing sentenceBert in this way is abrupt, more reliable on semantic evaluation?}
After post-processing, the performance in terms of $\mathrm{FENSE}_\mathrm{w/o \, penalty}$ almost remains unchanged. 
This demonstrates that post-processing reduces grammatical errors in captions without influencing semantic information.
%\ZYX{The reinforcement learning model that is prone to grammatical errors achieves significant improvement.
%Grammatical errors are much less in the baseline model, so the increase is small.}
The RL-trained model is more prone to generate grammatical errors in captions. Therefore applying error correction on the RL-trained model output achieves significant improvement. In contrast, there are much fewer grammatical errors in the baseline model output, especially on AudioCaps. The performance improvement brought by error correction is thus small.

Other metrics do not change significantly since they are not sensitive to grammatical errors~\cite{zhou2022can}.
It is worth mentioning that the CIDEr score of RL-trained models decreases after post-processing.
By RL training, high-frequency word redundancy and repetition are added to obtain higher scores while our proposed post-processing corrects some of these errors.
Therefore, the CIDEr score drops a little while the semantic content remains unaffected. 

\section{Conclusion}
In this paper, we propose a neural network-based approach to reduce false-repetition errors in AAC outputs.
We transform error correction to a sequence labeling task by recognizing words to be deleted in a caption.
Reference captions in AAC datasets are taken as clean sentences.
We corrupt clean sentences by manually-defined rules to get grammatically-erroneous sentences so that parallel ``corrupted-clean'' sentence pairs are generated.
A BiLSTM is trained on the synthetic parallel dataset, achieving high classification accuracy.
By applying the error corrector on AAC outputs, significant improvement is achieved in terms of fluency metrics, while the semantic content remains unaffected.


\bibliographystyle{splncs04}
\bibliography{ref}
%

% \printbibliography{}

\end{document}